\pdfoutput=1
\documentclass[prd,aps,nofootinbib,twocolumn,superscriptaddress,
preprintnumbers,balancelastpage,longbibliography]{revtex4-1}
\usepackage{ae,aecompl}
\usepackage{graphicx}	
\usepackage{amsmath}	
\usepackage{graphicx}
\usepackage{dcolumn}
\usepackage{bm}
\usepackage{graphics}
\usepackage{afterpage}
\usepackage{float}
\usepackage{subfigure}
\usepackage{rotating}
\usepackage{multirow}
\usepackage{tabularx}
\usepackage{booktabs}
\usepackage{multirow}
\usepackage{fancyhdr}
\usepackage{hyperref}
\usepackage[utf8]{inputenc}
\usepackage{theorem}
\usepackage{moreverb}
\usepackage{euscript}
\usepackage{psfrag}
\usepackage{slashed}
\usepackage{mathtools}
\usepackage{makecell}
\usepackage[flushleft]{threeparttable}
\usepackage[english]{babel}
\usepackage{bm}
\usepackage[mathlines]{lineno}
\usepackage{color}
\definecolor{rossoCP3}{cmyk}{0,.88,.77,.40}
\definecolor{darkred}{rgb}{0.6,0,0}
\definecolor{drkgrn}{RGB}{0, 51, 0}
\hypersetup{
     colorlinks   = true,
     citecolor    = blue,
     urlcolor     = magenta,
      linkcolor    = darkred
}

\usepackage{listings}
\usepackage{color,xcolor}

\newcommand{\be}{\begin{equation}}

\newcommand{\ee}{\end{equation}}
\newcommand{\bea}{\begin{eqnarray}}
\newcommand{\eea}{\end{eqnarray}}

\newcolumntype{C}[1]{>{\centering\let\newline\\\arraybackslash\hspace{0pt}}m{#1}}

\lstdefinestyle{python}{
  belowcaptionskip=1\baselineskip,
  breaklines=true,
  frame=L,
  xleftmargin=\parindent,
  language=Python,
  showstringspaces=false,
  basicstyle=\small\ttfamily,
  morekeywords={models, lambda, forms,True,False,None},
  keywordstyle=\bfseries\color{deepgreen!40!black},
  commentstyle=\itshape\color{gray},
  identifierstyle=\color{black},
  stringstyle=\color{deepred},
  rulecolor=\color{gray},
}

\begin{document}

\title{Reply to ``Comment on ``Damping of neutrino oscillations, 
decoherence and the lengths of neutrino wave packets"\,"}

\author{Evgeny Akhmedov}
\thanks{{\scriptsize Email}: \href{mailto:akhmedov@mpi-hd.mpg.de}
{akhmedov@mpi-hd.mpg.de}}
\affiliation{Max-Planck-Institut f\"{u}r Kernphysik, Saupfercheckweg 1, 69117 
Heidelberg, Germany}
\author{Alexei Y. Smirnov}
\thanks{{\scriptsize Email}: \href{mailto:smirnov@mpi-hd.mpg.de}
{smirnov@mpi-hd.mpg.de}}
\affiliation{Max-Planck-Institut f\"{u}r Kernphysik, Saupfercheckweg 1, 69117 
Heidelberg, Germany}

\date{\today}

\newcommand{\mk}[1]{{\bf #1}}
\newcommand{\om}[1]{\textcolor{red}{#1}}
\newcommand{\sh}[1]{\textcolor{blue}{#1}}

\begin{abstract}
In arXiv:2209.00561 \cite{jones}  
our treatment \cite{us} 
of effects of particles emitted together with neutrinos on 
neutrino wave packets is criticized on several grounds. We show here that this 
criticism is based on misinterpretation of our results and is invalid. 
Our conclusions and, in particular, the conclusion that neutrino wave 
packet separation effects are unobservable in reactor and neutrino source 
experiments, remain unchanged. 
\end{abstract}

\maketitle

In \cite{jones} it has been claimed that our treatment of damping of 
neutrino oscillations in sec.~3 of \cite{us} is inconsistent as it 
(i) suffers from causality violation, (ii) involves integration over 
non-orthogonal basis and (iii) inadequately considers the localization of 
the particles involved in neutrino production process. 
The author presumes that these points undermine our 
conclusion that decoherence by wave packet (WP) separation cannot be 
observed in reactor and neutrino source experiments. 
We demonstrate here that the first of the above claims is 
based on an incorrect interpretation of the space-time diagram serving to 
illustrate our treatment, the second claim criticizes a calculation we have 
never done and the third one stems from misinterpretation of the 
localization of the neutrino production process.

\section{Causality}
 
It is claimed in \cite{jones} that our analysis of damping of 
neutrino oscillations implies the possibility 
of superluminal signals and thus leads to causality violation. To illustrate 
this point, the author considers a two-stage thought experiment, in which 
initially neutrinos 
are produced in 
electron captures in a low-density gas, so that the interactions of the 
daughter nuclei with the surrounding particles of the medium can be 
neglected. This leads to relatively long neutrino 
WPs and no 
decoherence by WP separation observed in the detector. Then, at some 
``decision making time" $t_0$, one compresses the gas in the source with a 
piston, leading to strong localization of the daughter nucleus $N'$ and much 
shorter neutrino WPs; as a result, an observer at the detector position 
should see decoherence effects. However, according to the claim in 
\cite{jones}, this happens outside the future light cone with the origin at 
the point of $N'$ interaction soon after $t_0$, as the light signal from 
this point would reach the neutrino detector after the neutrino detection 
process has already been over. This would mean causality violation. 

The above argument is based on the incorrect space-time diagrams 
in Figs.~1 and 2 of \cite{jones} which do not correspond to our 
calculations. Our analysis in \cite{us} is based on a consideration of 
mean free times of the particles involved in the neutrino production 
process.  We have demonstrated that the production time is determined by 
the shortest among the mean free times $t_a$ of all the involved particles. 
In the cases considered in \cite{jones}, these are the mean free times of 
either electron (Fig.~1)%
\footnote{ 
Note that the light cone in Fig.~1 of \cite{jones} is plotted 
incorrectly: since the produced electron has in this case the shortest time 
of free propagation, the light cone should originate from the point of  
interaction with the surrounding atoms of the electron and not of $N'$.} 
or of the parent nucleus $N$ (both panels of Fig.~2), but not of $N'$. 
Thus, in all the diagrams of \cite{jones} the mean free time $t_{N'}$ of the 
daughter nucleus corresponds to the times when the neutrino production 
process has already been over. In such circumstances the interactions of 
$N'$ are irrelevant; they cannot (and do not) affect the outcome of the 
neutrino detection experiment. 

Consider now the situation when $t_{N'}<t_N$ 
(see Fig.~1 of this Reply). 
In this case the scattering of $N'$ {\em is} relevant and it determines the 
length $\sigma_x$ of the neutrino WP: $\sigma_x=ct_{N'}$. Simple 
geometrical considerations then show that, for baselines $L$ exceeding 
the coherence length $L_{\rm coh}=(2E^2/\Delta m^2)\sigma_x$ (which means 
that the WPs of the neutrino mass eigenstates $\nu_1$ and $\nu_2$ have 
separated by more than their length $\sigma_x$ before reaching the 
detector), the arrival of the slower neutrino mass eigenstate $\nu_2$ at 
the neutrino detector position will be inside the future light cone. 
This means that no causality violation arises in this case. 
\begin{figure}
\centering
\includegraphics[width=0.72\linewidth]{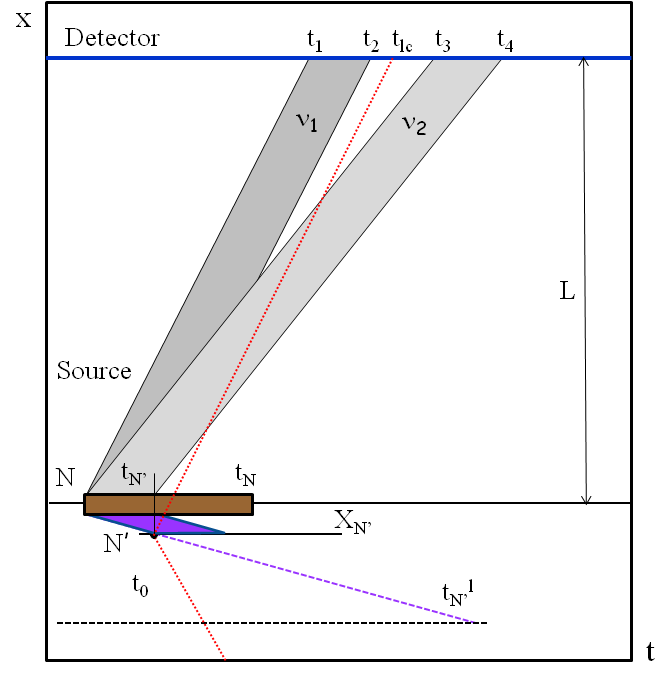}
\caption{
Space-time diagram of neutrino production, propagation and 
detection for neutrinos produced in $e$-capture in the case of compressed 
source gas (see text). Wave 
packets of the decaying nucleus $N$, daughter nucleus 
$N'$ and mass-eigenstate components of the emitted  
neutrino are represented by brown, violet and gray bands, respectively. 
$t_1$, $t_2$ and $t_3$, $t_4$ are the times of arrival at the detector of the 
``front'' and ``rear ends'' of the WPs of $\nu_1$ and $\nu_2$, respectively.   
Violet dashed line corresponds to the case of uncompressed gas in the source. 
Red dashed lines show the borders of the future light cone from the $N'$ 
interaction point.  The distance of this point from the source 
corresponds to the mean free path of $N'$. }
\label{fig:wp}
\end{figure}

Figure~2 of \cite{jones} actually corresponds to the situations when the 
duration $\tau$ of the individual processes of scattering of the daughter 
nucleus $N'$ on the surrounding atoms of the source rather than mean free 
time of $N'$ determines the length of the neutrino WP. (The author does not 
say this explicitly, but this follows from his figure). 
Note that such an approach is by itself fully legitimate, except that the 
time intervals $\tau$ have to be found rather than loosely assumed to be 
small or large, depending on the density of the system. This approach was 
mentioned in the end of sec. 3 of our paper \cite{us} as an alternative to 
the approach we have elaborated. If implemented, it could in principle give more accurate results for 
the lengths of the neutrino WPs than simple order-of-magnitude estimates we 
have obtained. The reason we did not pursue this approach in \cite{us} is 
that it would require the knowledge of the lengths of the WPs of all the 
particles participating in the scattering of $N'$ (including the WPs of the 
scattered states), which in turn would require taking into account their 
interactions with the surrounding atoms, and so on. Our approach, based on 
the consideration of mean free times, is in fact a shortcut allowing us to 
avoid considering this ladder of interactions. 

\section{Non-orthogonal bases}
It has been hypothesized in \cite{jones} that a possible reason for the 
alleged violation of causality in our approach was our use of WPs 
for describing propagation of daughter nuclei $N'$, and that WPs  constitute 
non-orthogonal bases. Using a non-orthogonal basis for summation over 
final states may indeed lead to wrong results. The point is, however, that 
we never performed such a summation, either explicitly or implicitly.  

It is argued in \cite{jones} that a summation over all the possible 
final states of the daughter nucleus $N'$ has to be performed, as the 
observer at the neutrino detector has no knowledge about these states. We disagree 
with this point. The only characteristic of $N'$ we use
is the total cross-section of its interaction with
particles of medium which determines 
the $N'$ mean free path.   
We have shown in \cite{us} that interactions of $N'$ 
with medium are only important for the formation of neutrino WPs when they 
happen on the time scales shorter than those of the interactions of the 
parent nucleus $N$. Therefore, they constitute an important part of the 
neutrino production process, 
and the production is not over until the scattering of $N'$ occurs. 
This scattering thus cannot be considered as happening ``after the 
neutrino production''. 

Notice that if one neglects the effects of $N'$ on neutrino WPs, 
this would greatly increase the lengths of these WP's, as was shown in 
sec.\ 3.1.1 of \cite{us}. This would only suppress decoherence effects 
and make their observation even less feasible. 

In \cite{jones} several issues related to the entanglement 
of the particles produced at neutrino detection and its connections 
with the EPR paradox were discussed. These are interesting points 
that deserve separate consideration. However, they are not directly 
related to the issues discussed in our paper \cite{us}.

\section{Localization due to nucleon interactions within the nucleus}
In \cite{us} we assumed that the localization of the atoms of neutrino 
source is caused by their scattering on surrounding atoms.
The author of \cite{jones} mentions as an alternative to this 
the localization of the decaying nucleon with respect to the other  
nucleons of the same parent nucleus. This would lead to much shorter 
neutrino WPs. 
However, such a localization through inter-nucleon interactions does 
not say anything about the absolute localization of the nucleus as a whole, 
and obviously only the latter is relevant for the formation of the neutrino WP. 
One might argue that the localization of the atom containing the unstable 
nucleus by its scattering on another atom does not determine the
absolute position of such an atomic pair in space either
\cite{jones2} (note that in \cite{us} we considered localization through
atom-atom scattering).
Let us clarify this issue.

Consider a source atom in a given coordinate system. The coordinate 
of the atom can be established (with some uncertainty) through its 
observable interactions with the other parts of the system. Consider first 
the situation when the source atom is placed in an empty box of linear 
size $l$. Then the coordinate uncertainty of the atom and of its nucleus 
(as well as of the constituent nucleons of the nucleus) will be given by 
$l$, not by the size of the atom or by the radius of its nucleus. This is 
because there is no way to find out where exactly inside the box the atom 
is, and its localization proceeds only through its collisions with the 
walls of the box. If the box is filled with a gas, the uncertainty 
of the coordinate of the parent nucleus will be given by the mean 
free path of its atom, because its scattering on the atoms of the gas 
will produce measurable recoils and local gas density changes. No 
such recoils are produced by the interactions of the decaying nucleon 
with the other nucleons of the same nucleus because the daughter 
nucleus recoils as a whole. Thus, such inter-nucleon interactions have 
nothing to do with localization of the parent particle in the source.
 
\vspace*{1mm}
To conclude, the criticism of our results in \cite{jones} is invalid. 
We maintain the validity of our results and, in particular, stand by our 
conclusion that WP separation effects cannot be observed in reactor and 
source experiments. 

\vspace*{1mm}
{\em Acknowledgement}. We thank B.J.P.~Jones for sending us his paper 
\cite{jones} before posting it on the e-print archive and for 
interesting discussions.


\end{document}